\documentclass[sigplan,nonacm]{acmart}
\pdfoutput=1

\AtBeginDocument{%
  \providecommand\BibTeX{{%
    \normalfont B\kern-0.5em{\scshape i\kern-0.25em b}\kern-0.8em\TeX}}}

\usepackage{stfloats}
\usepackage{subcaption}
\usepackage{graphics}
\usepackage{makecell}
\usepackage{algorithm,algpseudocode}

\setcopyright{none}
\renewcommand\footnotetextcopyrightpermission[1]{}
\begin{document}

\title{Armada: A Robust Latency-Sensitive Edge Cloud in Heterogeneous Edge-Dense Environments}

\author{Lei Huang, Zhiying Liang, Nikhil Sreekumar, Cody Perakslis, Sumanth Kaushik Vishwanath, Abhishek Chandra, Jon Weissman}
\affiliation{%
  \institution{University of Minnesota}
  \streetaddress{Twin Cities}
  \city{Minneapolis}
  \state{Minnnesota}
  \country{USA}
  \postcode{55455}
}
\email{{huan1397,liang772,sreek012,perak005, kaush047, chandra}@umn.edu,jon@cs.umn.edu}

\begin{abstract}
    Edge computing has enabled a large set of emerging edge applications by exploiting data proximity and offloading latency-sensitive and computation-intensive workloads to nearby edge servers. However, supporting edge application users at scale in wide-area environments poses challenges due to limited point-of-presence edge sites and constrained elasticity. In this paper, we introduce Armada: a densely-distributed edge cloud infrastructure that explores the use of dedicated and volunteer resources to serve geo-distributed users in heterogeneous environments. We describe the lightweight Armada architecture and optimization techniques including performance-aware edge selection, auto-scaling and load balancing on the edge, fault tolerance, and in-situ data access. We evaluate Armada in both real-world volunteer environments and emulated platforms to show how common edge applications, namely real-time object detection and face recognition, can be easily deployed on Armada serving distributed users at scale with low latency.
\end{abstract}


\maketitle

    \section{Introduction}
Edge computing, a computing paradigm that brings computation closer to data sources and end-users, has enabled the deployment of emerging edge-native applications \cite{bib:edgeNativeApp, bib:edgeNativeApp1}. With 5G accelerating the first network hop and rapid rollout of public edge infrastructure, edge computing is starting to play a significant role in the computing landscape \cite{bib:landscape}.

The emerging edge-native applications, including AR/VR, cognitive assistance, autonomous vehicles, are latency-sensitive and compute-intensive. Offloading workload from devices to powerful edge servers that can run complex machine learning algorithms is necessary to resolve the device-side limitation. The demand for these applications will increase rapidly and require the edge to be highly available and scalable. However, elasticity is a well-known limitation of edge resources \cite{bib:elasticity}. A burst of incoming workload can easily overwhelm an edge site causing service performance degradation. Furthermore, widely geo-distributed users require wide edge availability with full coverage of geographical locations to provide low-latency edge access. These requirements cannot be satisfied by single providers with limited point-of-presence and capacity in today’s edge infrastructure deployments \cite{bib:awsLocalZones, bib:awsWaveLength, bib:azureEdgeZones, bib:googleCloudEdge}.

Edge platforms that exploit edge resources from multiple providers have been proposed in both industry \cite{bib:mutable, bib:edjx, bib:mobileEdgeX}, and academia \cite{bib:edgenet} to enlarge the edge coverage. However, they are built on top of dedicated resources with a \textit{sparsely-distributed resource model}: users from a certain geographic location only have one or few nearby edge options which can provide a low-latency response. Overload can easily happen since dedicated resources are physically limited and lack scaling capabilities. With the advent of powerful personal computers and devices, we believe the necessary compute power is already closer to the users. Volunteer-based underused personal devices can be organized and coordinated at scale to resolve resource limitations on the edge. In this paper, we introduce Armada, a robust latency-sensitive edge cloud that explores the use of both dedicated and volunteer resources to support low-latency computation offloading. 

Armada uses a \textit{densely-distributed resource model}: users from a certain geographic location can have multiple nearby options to offload computations. Specifically, we explore the following challenges:

    \begin{itemize}
        \item How to select edge nodes to obtain low end-to-end latency in heterogeneous environments?
        \item How to achieve edge scalability with multiple loosely-coupled and resource-constrained edge nodes?
        \item How to guarantee continuous service in volunteer environments with high node churn and failure rate?
        \item How to minimize latency overhead for data persistence and consistency on edge?
    \end{itemize}

Armada implements auto-scaling service deployment mechanisms based on real-time user demand and distribution, 
and uses a user-side performance probing strategy as a key idea to guide service selection and load balancing among multiple edge nodes. 
The service deployment mechanisms incorporate several factors that affect performance, including user/data geo-location, edge server load, and network latency.
User-side probing employs multiple, flexibly maintained client-to-edge connections 
that provide fault tolerance by enabling immediate connection switch to alternate edge nodes upon node failure.
In addition, we introduce an edge-native storage layer to support low-latency data access when data and processing states cannot persist locally on volatile compute resources.

In this paper, we focus on the system and implementation aspects of Armada. We show how real-time inference, a common latency-sensitive and computation-intensive application category, can be easily deployed on Armada and serve geo-distributed users with low latency. Then we take a closer look at system scalability, fault tolerance and data access performance in both real-world volunteer environments and emulation environments. The evaluation shows that Armada achieves a 33\% - 52\% reduction in average user end-to-end latency with high concurrent demand compared to locality-based and dedicated-resources-only approaches.
    \section{Armada Overview}
In this section, we describe the heterogeneous edge-dense environment and give an overview of Armada design goals and system architecture. Then we discuss the application type that Armada supports.

    \subsection{Heterogeneous Edge-Dense Environment}
\begin{figure}[t]
                \centering
                \captionsetup{justification=centering}
                \includegraphics[scale=0.47]{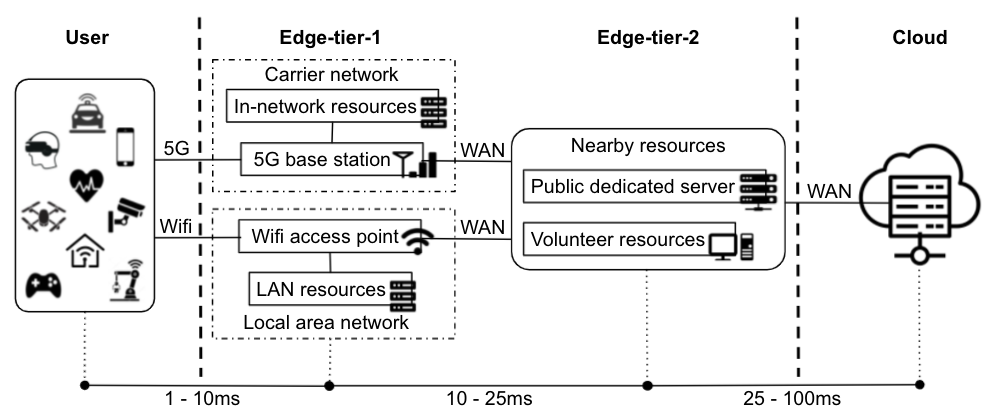}
                \vspace{-3mm}
                \caption{RTT latency in heterogeneous edge-dense environment}
                \vspace{-3mm}
                \label{fig:networking}
            \end{figure}

        Logical proximity, defined as low-latency high-bandwidth communication channels between edge servers and users, is usually provided by a LAN, on-premise networking infrastructures, and increasingly 5G technologies. However, special-purpose networking and compute resources on the edge are highly constrained in availability and scalability. In Figure \ref{fig:networking}, we show that nearby general-purpose resources in heterogeneous WAN environments (Edge-tier-2) can also provide low-latency benefits when Edge-tier-1 resources are not available or overloaded. We include both dedicated local public servers and volatile volunteer resources in Edge-tier-2 to enlarge the edge presence. Therefore, the resource limitation on edge can be resolved with the help of abundant volunteer edge nodes densely distributed around users, namely \textit{edge-dense} environments.

        The heterogeneity of Edge-tier-2 resources is twofold. First, connections from users to edge servers in WAN environments are highly diverse in terms of local ISPs and underlying networking infrastructure. Based on how users connect to the network, the actual number of routing hops and latency performance to the same edge server can highly diverge. Second, accessible compute resources present in nearby areas come from multiple providers and individuals. The heterogeneous capacity and hardware can lead to different processing performance, which is on the critical path of user requests and thus affects the end-to-end latency. Volunteer resources will amplify such heterogeneity by introducing more edge access points and increasing the system entropy.

    \subsection{Design Goals}
        Armada is designed with the following goals in mind:
        \begin{itemize}
            \item \textbf{Support for low-latency computation offloading at scale with densely distributed edge resources:} While one edge server is limited by its capacity, many loosely coupled but densely distributed edge nodes can coordinate with each other to provision nearby users at scale. Armada is designed to manage resource-constrained but abundantly distributed edge nodes to support scalable low-latency computation offloading. As a result, applications deployed on Armada are able to automatically scale and obtain more resources in a specific region if more users are present.
            \item \textbf{Locality-based service deployment:} Service deployment should be based on fine-grained geographical specifications to reduce networking latency. Multiple replicas \footnote{We use the term service replica and task interchangeably in this paper.} of the service should be deployed on different edge nodes to guarantee edge availability and capacity in specified regions. Changes to currently active users should also dynamically guide the service placement to fit the real-time user distribution. Furthermore, new service deployment should be optimized for short startup time to start serving users in a timely manner.
            \item \textbf{Performance-aware service selection in heterogeneous environments:} Geographical proximity is not strictly equivalent to low RTT latency. Multiple factors together determine the edge performance including network/compute resource heterogeneity and availability. Given a list of nearby edge nodes running replicas of the application service, Armada should identify the best-performing edge access point for each user to offload the computation. This edge selection process should also handle the load balancing for all users to achieve overall lower latency.
            \item \textbf{Ease of use:} Armada interfaces should be easy to use for both application developers and resource contributors. In particular, developers should use Armada SDK with minimum code modifications to their applications for deployment. Moreover, resource contributors should be able to register their nodes quickly with lightweight components and isolated runtime.
            \item \textbf{Fault tolerance:} Armada must ensure the fault tolerance for Armada users in the presence of high node churn due to volatile, unreliable and unpredictable volunteer resources. Armada users must be guaranteed continuous service and experience zero downtime upon node failure or node leaving.
            \item \textbf{In-situ edge storage:} Armada should provide a native storage layer on the edge \cite{bib:edgeNativeStore} to support low-latency data access. The storage layer should be reliable and independent from the volatile compute layer to persist the data for stateful and data-intensive applications. Also, flexible duplication and consistency policies should be supported for different application requirements.
        \end{itemize}

    \subsection{Armada Architecture}

        \begin{figure}[b]
            \centering
            \vspace{-3mm}
            \includegraphics[scale=0.77]{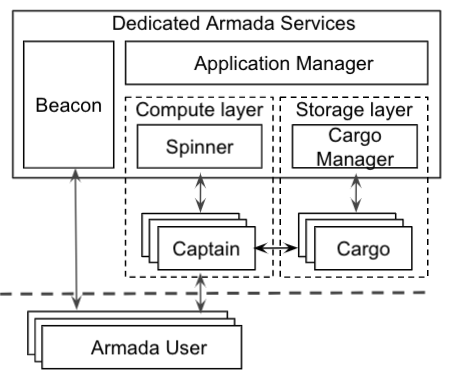}
            \vspace{-3mm}
            \caption{Armada system architecture}
            \vspace{-3mm}
            \label{fig:arch}
        \end{figure}

        Figure \ref{fig:arch} shows the Armada system architecture. Armada consists of geo-distributed nodes that donate their compute and/or storage resources, along with a set of global and central services hosted on dedicated, stable nodes. Both Armada system components and Armada-hosted applications are encapsulated in Docker containers for ease of use and fast deployment. Docker itself provides a lightweight, isolated runtime and abstractions over underlying resources for edge nodes, which is a good option for shipping the code easily to volunteer-based heterogeneous environments. Armada resources and services together constitute the following major components (described in Section \ref{components}):

        \begin{itemize}

            \item \textbf{Beacon:} Beacon is the global entry point for all interactions with Armada central services. It will forward requests to corresponding handler components, including application deployment requests, user connection requests and resource registration requests.

            \item \textbf{Application Manager:} Application manager maintains the states of submitted applications in Armada and manages the application lifecycle. It globally controls, operates, and monitors all application tasks running on different edge nodes, and processes initial user connecting requests. It also handles auto-scaling based on real-time user demand.

            \item \textbf{Compute Layer:} Compute layer manages dedicated and volunteer compute resources in Armada. It includes Spinner, the compute resource manager and Captain, the compute node. The Spinner handles compute node registration, health check and resource allocation for task deployment requests sent by the Application manager. The Captain manages the local heterogeneous resources through the Docker engine API and processes user workloads.

            \item \textbf{Storage Layer:} Storage layer manages dedicated and volunteer storage resources in Armada. It includes Cargo manager, the storage resource manager and Cargo, the storage node. The Cargo manager handles storage node registration, health check, maintains metadata and executes storage policies for data-dependent applications. The Cargo manages the local heterogeneous storage resources using the Docker volume and persists data on the edge supporting low-latency access for nearby users.

        \end{itemize}

    \subsection{Armada Applications} \label{sec:armadapp}

        Armada applications are long-running edge services using Armada resources for low-latency computation offloading. It includes a server-side program submitted to Armada for application-specific processing, and a client-side program used by application users to discover the service and offload computations. Armada deploys multiple replicas of the server-side program (tasks) to guarantee availability and scalability. Moreover, the client-side program uses Armada SDK to help application users locate the nearby service access points and establish direct communication channels. In Armada, we focus on the scenario where application users are co-located with the processing data, such as AR users sending out video streams for real-time processing. However, we also support external data upload from other data sources to the Armada storage layer, providing low-latency data access for running services.

        In Armada, volunteer resources are assumed to be unstable, volatile, and dynamic, with high node churn in heterogeneous environments. The guarantee on immediate recovery and continuous services upon node failure requires that application clients immediately switch connections to other service replicas and continue processing without waiting for failed node recovery. Therefore, no hard states or dependencies of the users are allowed to be maintained on the server-side for Armada applications. Application developers should either modify the application to maintain hard states and execution contexts on the client-side or use the Armada storage layer through Armada storage SDK to persist the data with minimized latency overhead.
    \section{Armada System Components} \label{components}
    
    \subsection{Beacon}
        Beacon is the entry point of contact for all initial interactions with Armada. It exposes interfaces for application developers to deploy edge services and monitor service status, application users to query service access points, and resource contributors to register edge nodes. Requests with different purposes will be forwarded to different handler services i.e., Application manager, Spinner and Cargo Manager, for further processing. Beacon provides the central public access point for different entities to establish initial connections with Armada components.
    
    \subsection{Application Manager} \label{AM}
        Application manager (AM) handles service deployment requests from application developers and service discovery requests from application users. AM also monitors the user demand and user distribution to make service auto-scaling decisions. Each service in Armada contains multiple replicas, namely tasks, deployed on distributed edge nodes. AM globally controls and monitors all replicas of the service through task-oriented APIs exposed by the compute layer (Section \ref{computeLayer}). In this way, Armada decouples the application-level management from the underlying edge resources layer. Three major modules of AM are described as follows.

            \begin{table}[b]
                \centering
                \captionsetup{justification=centering}
                \vspace{-3mm}
                \scalebox{0.97}{
                    \begin{tabular}{p{2cm} p{5.9cm}}
                        \hline
                        Parameter & Description \\
                        \hline
                        Image & Docker image for the application service \\
                        Compute\_Req & Compute resource requirements \\
                        Sched\_Policy* & Optional customized scheduling policy \\
                        Location & Coordinate(s) for expected user distribution \\
                        Need\_Storage & If persistent edge storage is required \\
                        Storage\_Req* & Storage requirements: capacity, consistency policy and data source \\
                        \hline
                    \end{tabular}
                }
                \vspace{1mm}
                \caption{Service deployment interface \\ (* denotes optional parameters)}
                \vspace{-6mm}
                \label{tab:serviceDeployment}
            \end{table}

        \textbf{Service deployment}. Initial service deployment request includes parameters shown in Table \ref{tab:serviceDeployment}. Service deployers only need to specify the resources required per replica without worrying about the number of replicas and replica distributions. AM initially deploys a minimum of three replicas to guarantee fault tolerance through the Spinner task deployment API. Then more replicas will be automatically spawned based on actual user demand and distribution (discussed later in auto-scaling). For all deployed tasks, AM periodically requests the underlying resource layer to collect real-time updates including running status, current load and resource utilization. If the \textit{Need\_Storage} field is true, AM will send storage resource requirements to the Cargo manager (Section \ref{storageLayer}) to allocate persistent edge storage capacity associated with the service.

        \textbf{Service discovery and selection}. AM maintains the metadata and states of all deployed service replicas. Application users need to query AM for nearby access points before establishing direct communication channels. However, the networking performance is nondeterministic in heterogeneous wide-area environments, and different hardware leads to different processing speeds. In addition, non-Armada networking traffic and workloads are unpredictable in practical volunteer environments, which will also cause performance fluctuation at random periods. There are no unified criteria to address all the above heterogeneities and system dynamics at the same time.

        We argue that periodic end-to-end latency probing is the only effective way to identify the best-performing edge node in real-time deterministically. In Armada, we propose a \textit{2-step} approach for application clients to select low-latency service access points accurately. AM implements the first step of this approach by generating the service \textit{candidate list}, and application clients finish the second step by performing the probing tests and making final decisions (Section \ref{armadaUser}).

        \begin{algorithm}[h]
          \footnotesize
          \caption{Service Selection Step-1}
          \label{alg:serviceSelection}
          \begin{algorithmic}[1]
            \Require{$Loc, NetType, \dots $}
            \Ensure{$CandidateList$}
            \Function{ServiceSelect}{$Loc, NetType, \dots$}
              \State {$LocalServices$ $\gets$ {$geoProximitySearch(Loc)$}}
                \For{$i \gets 1$ to $LocalServices.len()$}
                    \State {$EdgeNetType$ $\gets$ {$LocalServices[i].NetType$}}
                    \State {$Loc$}\parbox[t]{.9\linewidth}{%
                    {$LocalServices[i].Score \gets$} \\
                    {$LocalServices[i].Resources * weight1$} + \\
                    {$netAffiliation(EdgeNetType, NetType) * weight2$} + \\
                    {$\dots$}}
                \EndFor
                \State {$CandidateList$ $\gets$ $TopNSort(TopN, LocalServices)$ }
                \State \Return {$CandidateList$}
            \EndFunction
          \end{algorithmic}
        \end{algorithm}

        Algorithm \ref{alg:serviceSelection} shows how to generate the service \textit{candidate list} using the user information as input. The \textit{candidate list} is a small subset of service replicas that are likely to provide low latency responses for specific users. The considered factors include geo-proximity, resource utilization of the service replica (to detect overload), and the optionally-specified network affiliation between edge nodes and users. In \textit{geoProximitySearch()}, we apply GeoHash \cite{balkic2012geohash} with less precision to identify a wider-range geographical area, so relatively far-away edge nodes will be evaluated in the same way as closer edge nodes to avoid excluding better-performing options from the \textit{candidate list} in heterogeneous environments. \textit{TopN} (line 7) is the length of the \textit{candidate list}. Larger \textit{TopN} value leads to higher accuracy but also higher overhead during the performance probing step. We use the \textit{TopN} of 3 to have moderate overhead and enough accuracy.

        \textbf{Service auto-scaling}. AM handles the auto-scaling of the service based on the real-time user demand and distribution. The initial three service replicas are deployed in expected locations (Table \ref{tab:serviceDeployment}) without having actual users connected. When users join, AM will asynchronously associate user locations with new task deployment requests sent to the Spinner. Then, the Spinner scheduler will try to incrementally allocate more edge resources in specified locations to deliver better edge performance. With the help of Spinner scheduling policies (Section \ref{computeLayer}), AM auto-scaling requests can adapt to both higher user demand and wider user distribution by deploying more replicas in overloaded locations and spawning replicas in new locations.

        In Armada, scalability is achieved at both service deployment and user service selection levels to better allocate edge resources and to balance user workloads, achieving higher average performance.
    
    \subsection{Armada Compute Layer} \label{computeLayer}
        Armada compute layer manages dedicated and volunteer compute resources to execute latency-sensitive and computation-intensive edge services. It contains Spinner, the compute resource manager, and Captains, the geo-distributed edge compute nodes in Armada.

        \subsubsection{Spinner} \label{computeLayerSpinner}
        
            \begin{table}[b]
                \centering
                \vspace{-3mm}
                \scalebox{0.93}{
                    \begin{tabular}{p{2.1cm} p{2.4cm} p{3.6cm}} 
                        \hline
                        Interface & Input/Output & Description \\
                        \hline
                        Task\_Deploy & Task\_Metadata/ Status, Task\_ID & Application manager sends a task deployment request to Spinner. \\
                         
                        Task\_Status & Task\_ID/ Task\_Status & Application manager queries the runtime status of the task. \\
                        
                        Task\_Cancel & Task\_ID/Status & Application manager notifies Spinner to remove a task.  \\
                        
                        Captain\_Join & Node\_Metadata/ Status & A new Captain registers itself into the system. \\ 
                        
                        Captain\_Update & Captain\_Updates/ \_ & Captain sends heartbeats to Spinner reporting status updates. \\
                        
                        New\_Policy & Schedule\_Policy/ Status & Register a new scheduling policy. \\
                        \hline
                    \end{tabular}
                }
                \vspace{1mm}
                \caption{Spinner interfaces}
                \vspace{-3mm}
                \label{table:spinner-api}
            \end{table}
            
            Spinner manages edge compute resources in Armada, and runs the Armada scheduler that allocates edge resources and deploys tasks. Table \ref{table:spinner-api} shows Spinner interfaces, including task-oriented APIs for Application mananger to operate on tasks and APIs for Captains to register and report status. Spinner acts as the bridge between Armada applications and underlying edge compute resources.

            Spinner handles the \textit{Task\_Deploy} request through the Armada scheduler. Given the task image, resource requirements, target location, and optional custom scheduling policies, Armada scheduler uses a series of node filters followed by sorting policies to select edge nodes in heterogeneous environments effectively. We consider four types of policies:

            \begin{itemize}
                \item \textbf{Locality-based.} Geo-proximity filter is the fundamental policy to identify nearby edge nodes. Based on the density of edge nodes at target locations, the proximity range can be dynamically modified to limit the number of selected edge nodes.

                \item \textbf{Resource-aware.} Spinner monitors resource utilization (CPU and memory availability) of all edge nodes. Resource-aware sorting policy sorts the edge nodes based on the required compute power and actual availability.
                
                \item \textbf{Docker-aware.} The startup time of docker containers \cite{zheng2018wharf} can cause a high delay during the auto-scaling process when new service replicas need to be deployed very fast. Docker image layers with the same digest ID can be reused to reduce the downloading time of new images \cite{fu2020fast}. We use Docker-aware sorting policy to identify edge nodes that are faster to deploy tasks based on identical docker layers. 

                \item \textbf{Customized.} Application deployers can define custom filter and sorting policies to guide service scheduling. For example, network types and dedicated/volunteer resource preferences can be specified to sort or filter edge nodes. Data-dependent workloads can also specify policies to use data sources to guide node selection.
                
            \end{itemize}
            
            Filter policies are used sequentially to remove unqualified Captains, while all sorting policies are used collectively to determine the final sorting order. Each sorting policy is subject to a weight, defined as how significantly this policy affects the latency performance. The weighted score decides the final selected Captain for each \textit{Task\_Depoly} request. Note that Spinner also notifies un-selected Captains to prefetch the task images if possible to accelerate future task deployment by reducing the image downloading time.
        
        \subsubsection{Captain}
            Captain\footnote{Captain represents both the edge compute node and the controller container running in the node.} is an edge compute node in Armada. It listens to task operation instructions from Spinner, manages container lifecycle locally through Docker engine APIs, and discovers nearby edge storage capacity for data-related tasks using Cargo manager (Section \ref{storageLayer}). Captain isolates Armada runtime from the host environments and exposes edge services for direct connections with nearby users. Captain also reports local resource utilization, task running status and image repository information periodically to Spinner.

\subsection{Armada Storage Layer} \label{storageLayer}
    Armada storage layer maintains dedicated and volunteer storage resources in Armada. It enables edge services and applications to persist data on the edge with low-latency access. Armada storage layer consists of two components: Cargo Manager, the storage resource manager, and Cargos, the geo-distributed storage nodes.
    
    \subsubsection{Cargo Manager} 
    Cargo manager manages edge storage resources in Armada. Table \ref{tab:CMAPI} shows Cargo manager interfaces: for Cargos to join and report status, Application manager to allocate storage resources, and Captains to discover nearby data access points. Cargo manager also spawns data replicas to guarantee fault-tolerance and low-latency data access for geo-distributed services. Data persistence is achieved on edge with redundant data replicas and flexible data consistency policies. The three main modules of Cargo manager are described as follows:
    
    \begin{table}[h]
        \centering
        \scalebox{0.93}{
            \begin{tabular}{p{2cm} p{2.3cm} p{3.8cm}}
                \hline
                Interface & Input/Output & Description \\
                \hline
                Cargo\_Join & Cargo\_Metadata/ Status & A new Cargo registers itself into the system. \\
                
                Cargo\_Update & Cargo\_Updates/ \_ & Cargo sends heartbeats to Cargo manager reporting status updates. \\
                
                Store\_Register & Storage\_Req/ Status & Application manager registers storage capacity for an edge service. \\
                
                Cargo\_Discover & Captain\_Info/ Status & Captain queries nearby data access points \\
                \hline
            \end{tabular}
        }
        \vspace{1mm}
        \caption{Cargo manager interfaces}
        \vspace{-6mm}
        \label{tab:CMAPI}
    \end{table}      
    
    \textbf{Storage registration}: Application manager sends the \textit{Store\_Register} request to the Cargo manager during the service deployment phase (Section \ref{AM}) if the application requires persistent edge storage. The \textit{Store\_Register} request contains the service identifier, capacity requirement for each data replica, consistency policy, and the data source for original data uploading. We initially allocate resources and deploy three data replicas on three Cargos to guarantee availability and fault tolerance. The Cargo selection is based on locations and storage requirements given by the service deployment request.

    \textbf{Data access point selection}: Cargo manager maintains the metadata and states of all data replicas for an edge service. After the storage registration, Captain sends \textit{Cargo\_Discover} requests during the task deployment phase to help tasks find nearby data access points. A similar \textit{2-step} approach in service selection is applied to overcome the network heterogeneity and locate the best-performing data access point. First, a candidate list is generated by the Cargo manager based on the geo-proximity between the Captain and Cargos holding the data replicas. Optional factors like network affiliation can also be specified to help rank the candidates. Second, Captain performs the data access probing to identify the fast access point. The additional candidates in the list are used to handle fault tolerance through immediate connection switch upon Cargo failures.

    \textbf{Storage auto-scaling}: Initially a service is allocated three storage replicas. When more service replicas are spawned to satisfy higher user demand and wider user distribution, the storage layer should also adaptively scale to guarantee low-latency data access for geo-distributed service replicas. We employ the similar idea applied in the service auto-scaling process. When new service replicas are deployed, the Cargo manager asynchronously creates new data replicas on geo-proximate Cargos to the services. Since more replicas lead to higher resource usage and data consistency overhead, the Cargo manager collects the data access probing feedback from Captains to evaluate the need to spawn new data replicas carefully.

    \subsubsection{Cargo nodes} Cargo is an edge storage node in Armada. It handles data I/O operations and propagation of updates to replicas depending on the type of consistency. Each Cargo node is aware of at most three replica Cargo nodes corresponding to application data. The updates made to one Cargo node are propagated in a cascade manner to all the replicas if more data replicas are spawned to meet the user demands. Table \ref{tab:astore} describes the Armada storage SDK used by server-side application programs to interact with the storage layer. With Captains locating nearby data access points, Armada storage SDK helps tasks transparently communicate with nearby Cargos.
    
    \begin{table}[h]
        \centering
        \vspace{-1mm}
        \scalebox{0.93}{
            \begin{tabular}{p{1.8cm} p{2.2cm} p{4cm}}
                \hline
                Function & Input / Output & Description \\
                \hline
                Init\_Cargo & Cargo\_App\_ Metadata/ Status & Establish connection with a Cargo node\\
                
                Write & Write\_Data/ Write\_Status & Write data to the Cargo node\\
                
                Read & Read\_Data/ Read\_Status & Read data from Cargo node\\
                
                Close\_Cargo & \_/Status & Close connection to Cargo node after use \\
                \hline
            \end{tabular}
        }
        \vspace{1mm}
        \caption{Armada storage SDK}
        \vspace{-6mm}
        \label{tab:astore}
    \end{table}

    \section{Application Client} \label{armadaUser}
	Application client is the user-side program of Armada applications. It contains the application-specific logic and uses Armada client SDK to help application users locate the service access points. Application client plays an important role in coordinating with Armada system components to achieve latency-sensitive service selection, scalability and fault tolerance. We describe performance probing and multi-connection strategies which are core building blocks inside Armada client SDK, and discuss how they are applied to deliver Armada benefits.

	Performance probing, as discussed in Section \ref{AM}, is the second step in the service selection process. Application clients first obtain the service candidate list through the Beacon interface and then establish connections to each candidate for probing tests. The candidate with the lowest end-to-end latency is selected to start offloading the actual workload. More importantly, the \textit{2-step} service selection process is performed periodically and asynchronously in the background to adapt to system dynamics. If the selected node is overloaded or a closer node joins the system later, application clients can always identify the changes and switch to a better edge node if necessary. As a result, load balancing is automatically handled since overload can negatively affect the performance probing results. A far-away edge node can be selected if a closer node delivers worse performance due to overload. Therefore, latency-driven performance probing balances the load and improves edge scalability.

	Multi-connection strategy is used to achieve fault tolerance and guarantee continuous service. Each application client maintains multiple connections to different candidate edge nodes and uses this redundancy to prepare for potential server failures. Since all connections are already established and processing data is independent from the server (Section \ref{sec:armadapp}), no additional overhead is present to switch connections from the failed node to a working node. Candidate nodes obtained from the service selection process are already sorted by performance, therefore the second-best candidate is selected to maintain low-latency responses.

	Application developers develop the application client program using Armada client SDK to easily integrate above functionalities with minimum code modifications. We currently support the gRPC protocol in Golang and around 10 lines of code are added to apply the changes in our experiment applications.

    \section{Real-time Inference on Armada}
    We implement two real-time inference workloads to evaluate Armada performance. Real-time object detection and face recognition are critical building blocks in commonly used applications like augmented reality, cognitive assistance and security surveillance. They are both computational-intensive and latency-sensitive, which require offloading the computation to powerful servers and obtaining processing results in a timely manner. First, we use an object detection workload to demonstrate the workflow of the Armada computing layer. Second, the face recognition workload \cite{bib:goface} showcases the coordination between computing and storage layer when Armada application needs persistent edge storage.

    \subsection{Real-time Object Detection}
\begin{figure}[t]
                \centering
                \includegraphics[scale=0.62]{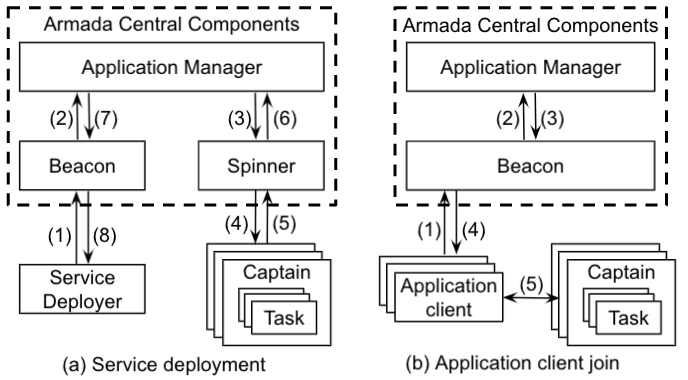}
                \vspace{-3mm}
                \caption{Object detection workflow in Armada}
                \label{fig:objDetect}
            \end{figure}

        Figure \ref{fig:objDetect} shows the workflow of real-time object detection in Armada. In the service deployment phase (Figure \ref{fig:objDetect} (a)) , service deployers first contact Beacon in step (1) to submit the application along with requirements to Armada. Application manager receives this request in step (2) and initiates three task deployment requests sent to Spinner in step (3). Spinner then calls the Armada scheduler to find available edge nodes and place the tasks in step (4). In the end, the tasks deployment status and service deployment status are updated back to the deployers in step (5) - (8).

        In Figure \ref{fig:objDetect} (b), When users request the object detection service in Armada, they need to first query the system for service access points in step (1) - (4), and then start sending the video frames for object detection in step (5). Note that \textit{TopN} number of connections are maintained using the \textit{candidate list} obtained from the service selection process.

            \begin{figure}[t]
                \vspace{-3mm}
                \centering
                \includegraphics[scale=0.57]{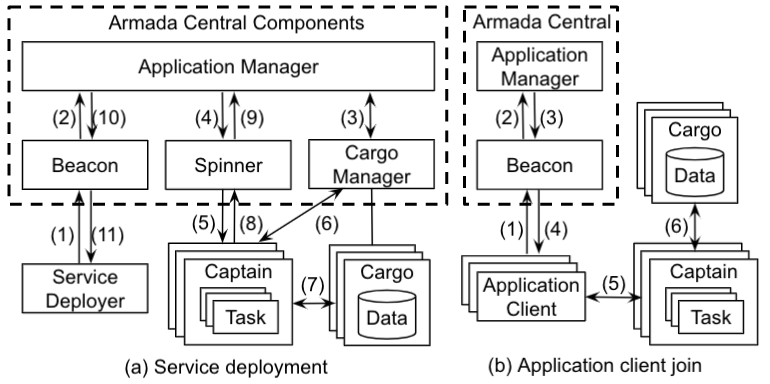}
                \vspace{-3mm}
                \caption{Face recognition workflow in Armada}
                \vspace{-3mm}
                \label{fig:faceRec}
            \end{figure}    

    \subsection{Face Recognition}
        \label{subsec:facerec}
        Figure \ref{fig:faceRec} shows the workflow of real-time face recognition in Armada. In the service deployment phase (Figure \ref{fig:faceRec}(a)), service deployers first submit the application along with requirements for both compute and storage resources (1) - (2). Then the Application manager contacts the Cargo manager to register the storage requirement of the service (3). The Cargo manager selects three Cargos and allocates the required storage resources for three data replicas. The three Cargos then use the specified data source to pull the initial pre-labeled face datasets used to recognize people during the real-time inference. In step (4) - (5), tasks are sent to the compute layer for deployments. To connect tasks with nearby data access points, Captains queries Cargo manager in step (6). Given the \textit{candidate list} of access points, tasks can directly interact with the selected data replicas in step (7) using Armada storage SDK. In the end, the task and service deployment status are updated back to the deployers in step (8) - (11).

        Figure \ref{fig:faceRec} (b) shows the workflow when face recognition clients request the service. In step (1) - (4), clients first query the system for service access points, and then start sending video frames for face recognition in step (5). For any detected faces during the processing, tasks query data replicas in Cargos for face recognition (6). The read requests send detected faces to Cargo searching for matched people, and the write requests insert new labeled faces into the persistent data store for future recognition.
    \section{Evaluation}

    We evaluate Armada in both real-world edge environments and emulation platforms in the cloud. The real-world experiment explores Armada performance in fine-grained small geographical areas (regions within a city). The emulation experiment explores wider-range geographical areas (regions across nearby cities). We first use a computation-only workload, object detection, to demonstrate Armada service selection, scalability and fault tolerance performance. Then we use a face recognition workload to explore the storage layer performance when the persistent store is required.

    \subsection{Experimental Setup}

        In Table \ref{tab:setup}, we show the underlying hardware used for both real-world and emulation experiments. Note that the third column shows the processing time per frame for real-time object detection application on these hardwares.

        \begin{table}[b]
            \begin{subtable}{0.47\textwidth}
                \vspace{-3mm}
                \centering
                \scalebox{0.923}{
                    \begin{tabular}{|p{0.85cm}|p{5.5cm} p{1.5cm}|}
                        \hline
                        Node & Processor & Processing \\
                        \hline
                        V1 & Intel® Core™ i7-9700, 8 cores & 24ms \\
                        V2 & Intel® Core™ i7-2720, 6 cores & 32ms \\
                        V3 & Intel® Core™ i9-8950HK, 6 cores & 31ms \\
                        V4 & Intel® Core™ i5-8250U, 4 cores & 45ms \\
                        V5 & Intel® Core™ i5-5250U, 2 cores & 49ms \\
                        D6 & Intel® Xeon® CPU E5-2620 v3, 24 cores & 30ms×4 \\
                        Cloud & t2.large, 4 cores & 34ms \\
                        \hline
                    \end{tabular}        
                }
                \caption{Real-world experiment}
                \label{tab:setup1}        
            \end{subtable}

            \begin{subtable}{0.47\textwidth}
            \centering
            \scalebox{0.94}{
                \begin{tabular}{|p{0.8cm}|p{2.5cm} p{2.5cm} p{1.5cm}|}
                    \hline
                    Node & Type & Location & Processing \\
                    \hline
                    A & t2.2xlarge, 8 cores & City\_A & 23ms \\
                    B & t2.large, 4 cores & City\_B & 34ms \\
                    C & t2.small, 2 cores & City\_C & 58ms \\
                    Cloud & t2.large, 4 cores & Cloud & 34ms \\
                    \hline
                \end{tabular}    
            }
            \caption{Emulation experiment}
            \label{tab:setup2}        
            \end{subtable}
            \caption{Hardware used and frame processing performance}
            \vspace{-3mm}
            \label{tab:setup}
        \end{table}

        \subsubsection{Real-world Environment} \label{ree}
            We set up the real-world experiment environment around our University campus. As shown in Table \ref{tab:setup} (a), A combination of both dedicated and volunteer resources is used. Volunteer nodes V1 - V5 are located within 5 miles of the campus, and a powerful University server D6 located on campus is considered the dedicated edge node.

\begin{figure}[t]
                \centering
                \includegraphics[scale=0.65]{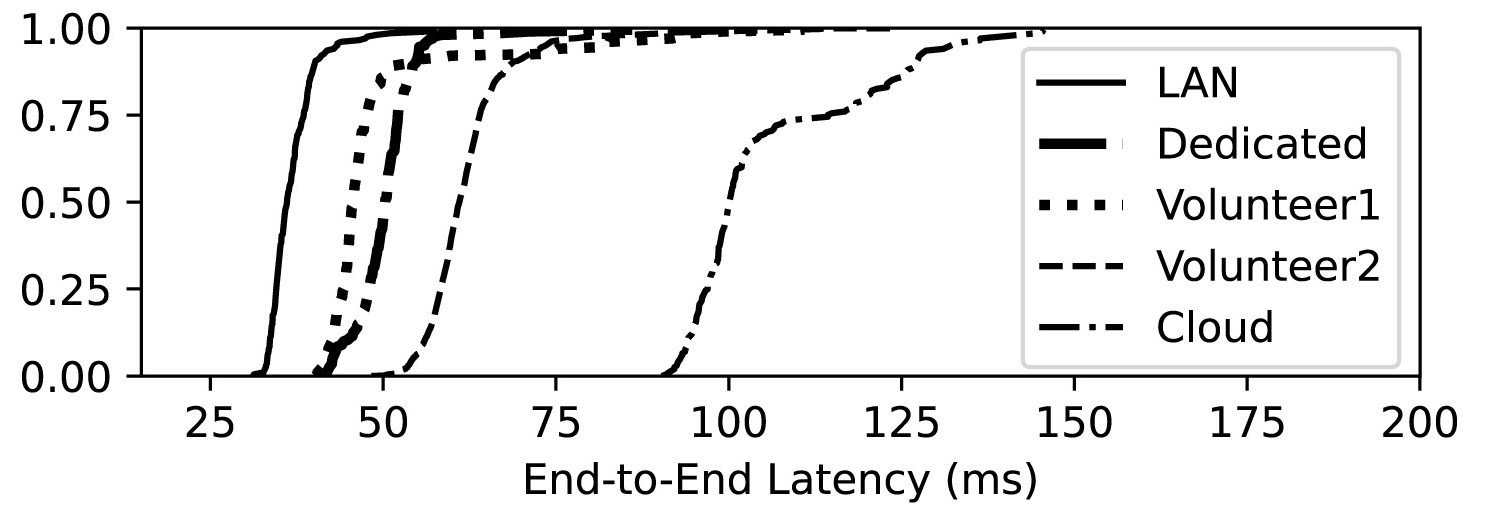}
                \vspace{-3mm}
                \caption{CDF of end-to-end latency for different servers}
                \vspace{-3mm}
                \label{fig:1-1-1}
            \end{figure}

            While the dedicated node has more compute power and better network connectivity, volunteer nodes are set up with heterogeneous compute and networking performance contributed by actual volunteers around the campus. The dedicated node D6 can hold four service replicas in parallel, with each of them processing the video at 30ms/frame. Figure \ref{fig:1-1-1} shows the benefits of exploiting volunteer resources from one user's perspective. Volunteer nodes can deliver similar or even better performance compared to the dedicated edge node.

        \subsubsection{Emulation Environment} \label{emulationSetup}
            Due to physical limits, we use the emulation environment to explore Armada performance on a wider geographical scale. We use the network emulation platform Netropy \cite{bib:netropy} in AWS to emulate WAN connectivity for three nearby cities City\_A, City\_B and City\_C that are about 100 - 150 miles away from each other. Three edge nodes A, B, C are located at three locations as shown in \ref{tab:setup} (b). 

        \subsubsection{Baselines}
            We use geo-proximity, dedicated-edge-only and cloud scenarios for comparisons with Armada.
            \begin{itemize}
                \item \textbf{Geo-proximity}: In the geo-proximity scenario, we force all users to connect to the closest edge node in a geographical location, a typical edge selection policy to identify the low-latency edge access point.
                
                \item \textbf{Dedicated-edge-only}: In the dedicated-edge-only scenario, we assume that only limited dedicated edge resources are available, which is common in today's edge infrastructure deployment. As shown in Table \ref{tab:setup} (a), we use one powerful dedicated node as compared to 5 resource-constrained volunteer nodes to maintain a reasonable ratio of the availability of dedicated and volunteer resources. We show the benefits of exploiting volunteer resources by comparing them with the dedicated-edge-only scenario.
                
                \item \textbf{Cloud}: We show the cloud performance as the baseline compared to other scenarios. We use the closest AWS service region US East to deploy the services and assume that the cloud has unlimited scalability with increasing user demand.
            \end{itemize}

    \subsection{Latency-Sensitive Service Selection}
        We set up three users C1, C2 and C3, in the real-world experiment. They are located around the campus with heterogeneous networking performance to different edge nodes. We also set up three users User\_A, User\_B and User\_C, in the emulation platform and configure them to be at the same locations as nodes A, B and C with corresponding real-world WAN networking performance. Table \ref{tab:selection} shows the pairwise end-to-end latency for object detection application. The bold underlined values refer to the selected service access point in Armada for each user.

        \begin{table}[h]
            \begin{subtable}{0.42\textwidth}
                \vspace{-1mm}
                \centering
                \scalebox{0.95}{
                    \begin{tabular}{|p{0.85cm}|p{0.5cm} p{0.5cm} p{0.5cm} p{0.5cm} p{0.5cm} p{0.5cm} p{0.85cm}|}
                        \hline
                        Client & V1 & V2 & V3 & V4 & V5 & D6 & Cloud\\
                        \hline
                        C1 & \underline{\textbf{38}} & 47 & 49 & 65 & 72 & 42 & 107 \\
                        C2 & 43 & \underline{\textbf{35}} & 56 & 58 & 61 & 45 & 102 \\
                        C3 & 49 & 50 & 45 & 59 & 71 & \underline{\textbf{42}} & 112 \\
                        \hline
                    \end{tabular}        
                }
                \caption{End-to-end latency (ms) in real-world environment}
                \label{tab:selection2}        
            \end{subtable}
            \begin{subtable}{0.42\textwidth}
                \centering
                \scalebox{0.95}{
                    \begin{tabular}{| p{1.7cm} | p{1cm} p{1cm} p{1cm} p{1cm} |}
                        \hline
                        Client & A & B & C & Cloud\\
                        \hline
                        User\_A & \underline{\textbf{31}} & 63 & 89 & 108 \\
                        User\_B & 63 & \underline{\textbf{47}} & 83 & 102 \\
                        User\_C & \underline{\textbf{51}} & 68 & 58 & 111 \\
                        \hline
                    \end{tabular}        
                }
                \caption{End-to-end latency (ms) in emulation environment}
                \label{tab:selection1}
            \end{subtable}
             \caption{Latency-sensitive service selection in Armada}
             \vspace{-5mm}
             \label{tab:selection}
        \end{table}
         \vspace{-2mm}

        Table \ref{tab:selection} (a) and (b) show that users in both real-world and emulation environments can identify the heterogeneity of the environment and select the best-performing node to offload the workload. In Table \ref{tab:selection} (b), User\_C can select a farther node A due to local resource limitation in node C.

    \subsection{Scalability and Load Balancing}
        We explore Armada's scalability performance over high user demand and wide user distribution. We evaluate the average end-to-end latency for the object detection application with a varying number of users and edge nodes.

        \subsubsection{Performance over increasing user demand}
\begin{figure}[!h]
                    \centering
                    \vspace{-2mm}
                    \includegraphics[scale=0.37]{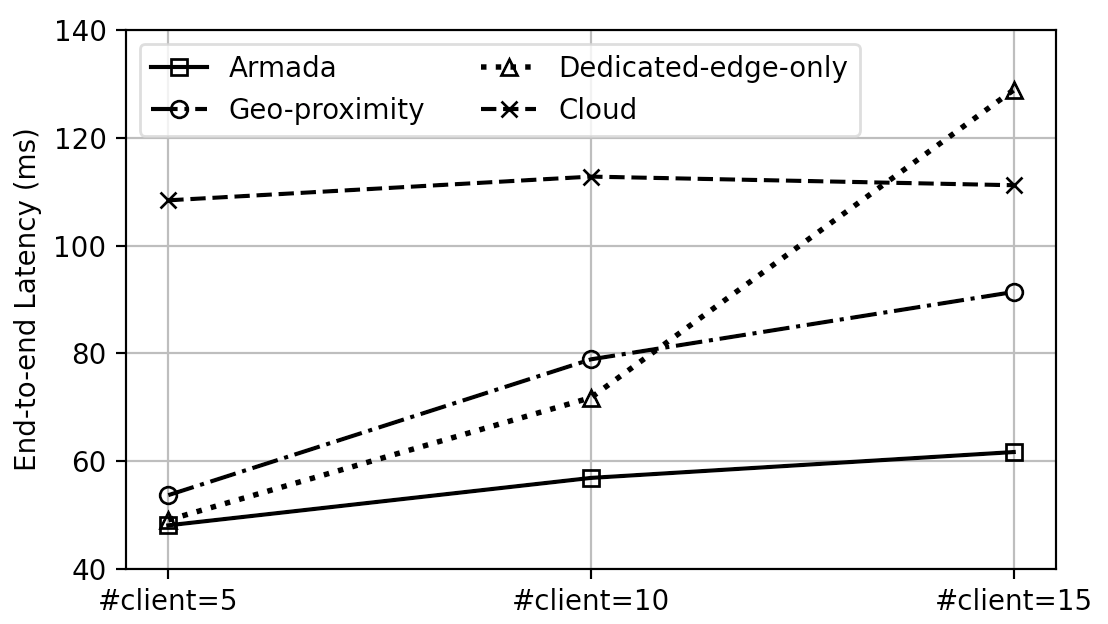}
                    \vspace{-3mm}
                    \caption{Performance over increasing user demand}
                    \vspace{-2mm}
                    \label{fig:3-1-2}
                \end{figure}

            We recruit 15 users around the campus (within 5 miles) with heterogeneous networks to play object detection clients in real-world experiments. With edge resources from five volunteer nodes and one dedicated node shown in Table \ref{tab:setup} (a), 15 users incrementally start requesting the service. We record the average end-to-end latency at three time slots when there are five, ten and 15 concurrent users. Figure \ref{fig:3-1-2} shows the user average performance using Armada as well as other baselines.

            Armada shows promising scalability performance: 33\% faster than the geo-proximity scenario and 52\% faster than the dedicated-edge-only scenario at \#client = 15 in our experimental setup. First, locality-based service selection ignores network heterogeneity and quickly leads to performance degradation caused by overload. Second, dedicated edge resources are limited in point-of-presence and elasticity. High concurrent user demand can easily overload an edge cite as shown in Figure \ref{fig:3-1-2}, where the dedicated-edge-only scenario is even worse than cloud performance at \#client = 15.

        \subsubsection{Performance over wide user distribution}
            In this emulation experiment, we explore Armada scalability and load balancing behaviors in wide area settings.

\begin{figure*}[t]
            \begin{center}
                \centering
                \includegraphics[scale=0.57]{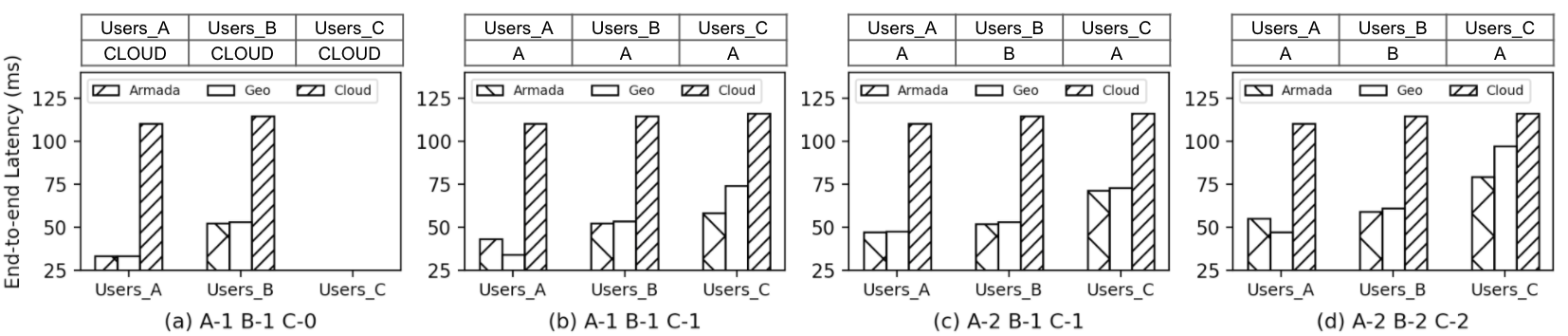}
                \vspace{-3mm}
                \caption{End-to-end latency: varying no. of users with fixed set of edge nodes. Each subfigure shows performance under different user distributions. The notation table tells the user edge selection results in Armada.}
                \vspace{-3mm}
                \label{fig:2-2-1}
            \end{center}
            \end{figure*}

            \textbf{Varying no. of users with a fixed set of edge nodes}: In Figure \ref{fig:2-2-1}, with static edge nodes A, B and C as described in \ref{tab:setup} (b), we incrementally add users to different cities and observe the average latency performance for users at each city. Each subfigure tells the user distribution and the notation table tells the user edge selection results in Armada. Figure \ref{fig:2-2-1} (a), as an example, has one user at City\_A, one user at City\_B and zero user at City\_C. The City\_A user selects node A and the City\_B user selects node B for processing. We also show the latency performance for locality-based edge selection and cloud as comparisons with Armada.

            Figure \ref{fig:2-2-1} (b) shows that the user at City\_C selects node A for processing since node A is more powerful and has better performance compared to local node C. Figure \ref{fig:2-2-1} (c) shows that when two local users are present at City\_A, the user at City\_C switches back to local node C since node A is fully loaded serving local users. Figure \ref{fig:2-2-1} (d) shows that when node C is already serving a local user, the second user selects the farther node A after performance probing comparisons. Note that the average performance for users at City\_A in Figure \ref{fig:2-2-1} (b) and (d) are worse than the locality-based approach because local node A serves more users from other cities.

\begin{figure*}[t]
            \vspace{-3mm}
            \begin{center}
                \centering
                \includegraphics[scale=0.57]{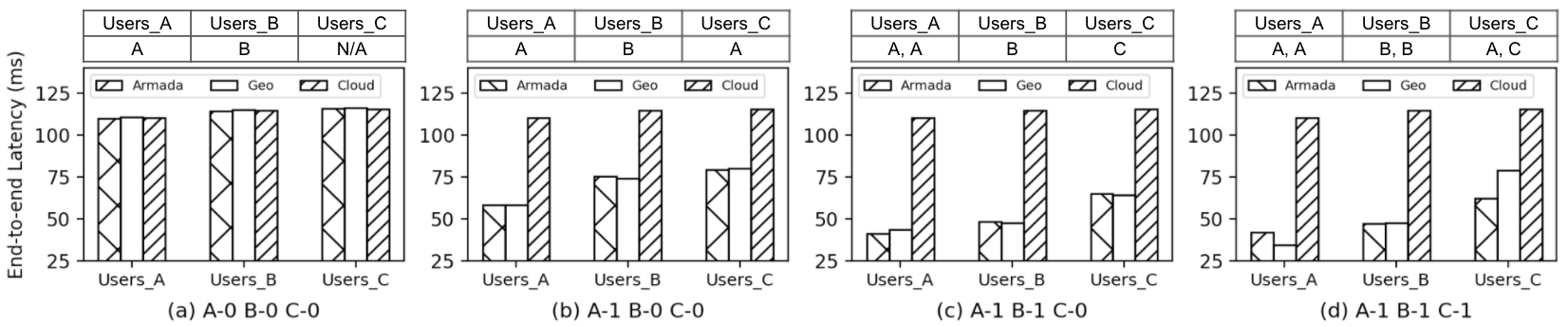}
                \vspace{-3mm}
                \caption{End-to-end latency: varying no. of edge nodes with fixed set of users. Each subfigure shows performance under different edge node distributions. The notation table tells the user edge selection results in Armada.}
                \label{fig:2-2-2}        
            \end{center}
            \vspace{-3mm}
            \end{figure*}

            \textbf{Varying no. of edge nodes with a fixed set of users}: In Figure \ref{fig:2-2-2}, with three static users at three cities, we incrementally add edge nodes to observe the user performance. Subfigure captains tell the edge node distribution in this case. Figure \ref{fig:2-2-2} (b) shows that a new node at City\_A improves the performance of all three users in different cities. Figure \ref{fig:2-2-2} (c) shows that a new node at City\_B further improves the performance of all three users. The user at City\_B switches to local node B and releases more resources in node A. Figure \ref{fig:2-2-2} (d) shows that a new node at City\_C does not affect the performance because the powerful node A delivers a better performance to the user at City\_C.

        \subsubsection{Fast auto-scaling and Captain registration}
\begin{figure}[H]
                \centering
                \vspace{-2mm}
                \includegraphics[scale=0.4]{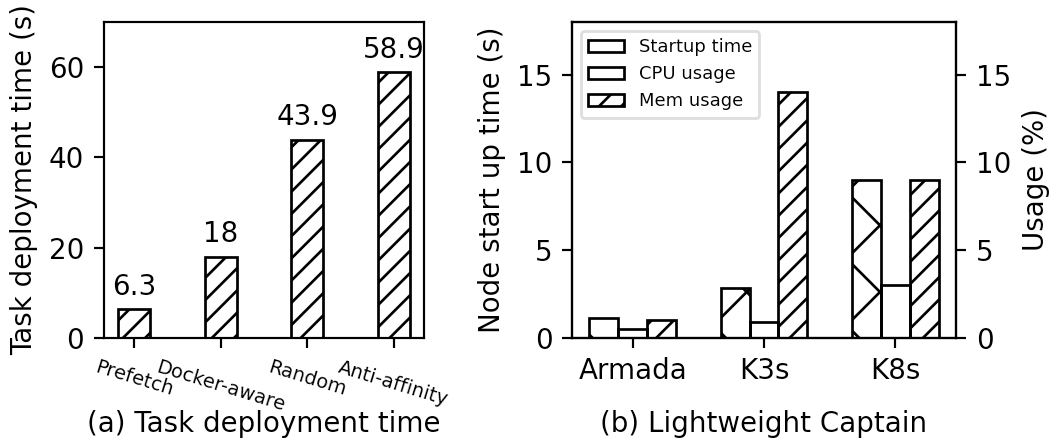}
                \vspace{1mm}
                \caption{Fast auto-scaling and Captain registration}
                \vspace{-3mm}
                \label{fig:lightweight}
            \end{figure}

            We also explore the task deployment speed during the service auto-scaling process. Figure \ref{fig:lightweight} (a) shows the average task deployment time based on different strategies. When multiple edge nodes satisfy the task deployment requirements, Armada uses image prefetch and Docker-aware policies discussed in Section \ref{computeLayerSpinner} to reduce the deployment time. As compared to random selection and anti-affinity selection \cite{bib:antiaffinity}, a common approach to avoid workload similarities, Armada implements faster task deployment.

            Armada has unlimited potential to expand with the help of volunteer nodes. In Figure \ref{fig:lightweight} (b), we measure the Captain registration time and resource usage during idle time to explore Captain lightweight characteristics. It shows that Captain is 57\% and 86\% faster than K3s  \cite{bib:k3s} and K8s \cite{brewer2015kubernetes} registration and has lower resource usage during idle time. Note that we only record the time used for node registration modules in K3s and K8s for fair comparisons.

    \subsection{Fault Tolerance}
        Armada uses the user-driven multi-connection strategy to guarantee continuous service over edge failures. We evaluate the Armada fault tolerance performance in the real-world experiment environment with the object detection workload.

\begin{figure}[H]
            \centering
            \vspace{-3mm}
            \includegraphics[scale=0.56]{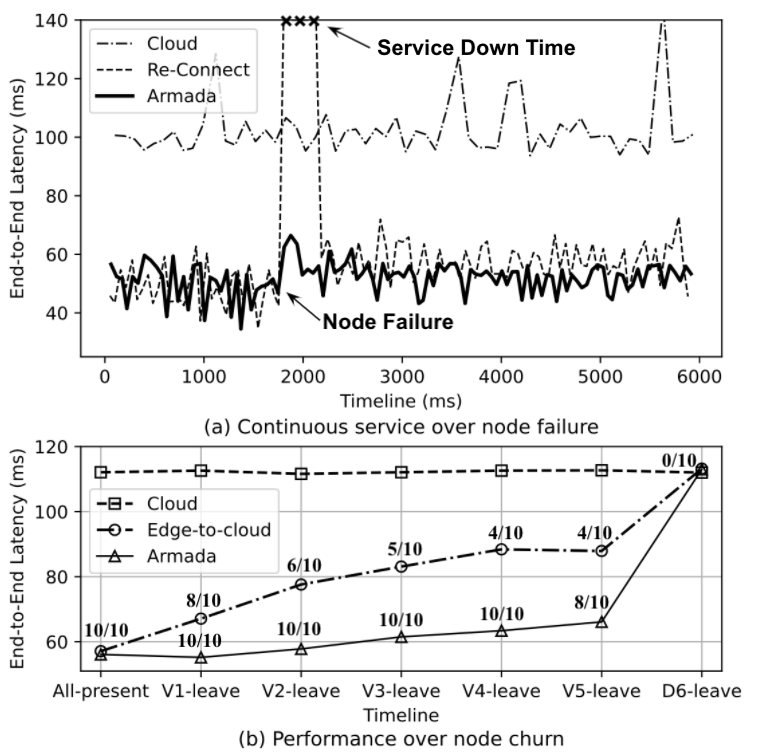}
            \vspace{-3mm}
            \caption{End-to-end latency over node churn. The ratios over data points show the number of users that are still connected to edge nodes.}
            \vspace{-3mm}
            \label{nodeChurn}
        \end{figure}

        Figure \ref{nodeChurn} (a) shows the end-to-end latency for continuous video frames from a single-user perspective. When the currently connected edge node suddenly fails or leaves the system, the Armada client can immediately switch to a backup node and prevent the service downtime compared to a server re-connect approach.

        In Figure \ref{nodeChurn} (b), we manually fail edge nodes one by one and observe the average end-to-end latency of ten static users after each failure. The service is always guaranteed to be continuous in this experiment. So, as comparisons, we develop an Edge-to-Cloud approach where the end-user can immediately switch to the cloud due to node failure. The value on top of each data point (say 8/10) shows the number of still connected users to the edge after each node failure. With all the edge nodes failing, both Edge-to-Cloud and Armada approaches show cloud performance at the end. However, Armada shows a lower average latency since the failed users switch to alternative edge nodes for low-latency processing.

    \subsection{Performance of Storage Layer}
        We use the face recognition workload to evaluate storage layer performance in the real-world experiment. In the following experiments, we focus on the communications between tasks and Cargos. Therefore we configure the \textit{TopN} to 1 to simplify the compute layer workflow. In this case, each application client only connects to one task.

        We explore the effects of the Cargo selection strategy, storage fault tolerance and different consistency policies. The same set of resources described in Table \ref{tab:setup1} (a) is used, with each one of them having 2GB persistent storage capacity. In addition, each data replica initially uploaded to Cargo contains 1000 labeled face descriptors \cite{bib:lfw} in the format of $<$ID (8 bytes), vector (128 * 8 bytes)$>$ pairs. We focus on three workloads for evaluation:

            \textbf{Read-only workload}: 1000 face images are used as the task input video frames for real-time recognition. The task processes each image, detects the face and generates a unique face descriptor. Then the task queries Cargo to find the matched descriptor along with the face ID. The read latency includes the time to connect to the Cargo and query processing. The tasks do not buffer labeled faces locally to explore the Armada storage layer performance thoroughly.
            
            \textbf{Write-only workload}: 1000 new face images are used as the task input video frames. We configure the task to detect faces and directly write new face descriptors with face IDs into the Cargo data replica. The write latency includes the time to connect to the Cargo and to perform the writing.
            
            \textbf{Read-followed-by-write workload}: 1000 new face images are used as the task input video frames. For each image, the task first sends a read request to query the Cargo and then writes the new face descriptor into the Cargo when the read request cannot recognize the face.

        \subsubsection{Cargo selection} \label{cargoSel}
            We explore the Cargo selection results using the read-only workload. Nodes V1, V2, D6 and Cloud are registered as four Cargos, and V3, V4 and V5 are used as Captains to run three face recognition tasks. We also configure three users co-located with three Captains for simplicity. Table \ref{tab:cargoSelect} shows the Cargo selection result and pairwise read latency. We can see that the Cargo selection strategy can identify the environmental heterogeneity and select the best-performing data access point for each data-dependent task.

            \begin{table}[!h]
                \centering
                \vspace{-1mm}
                \scalebox{0.91}{
                    \begin{tabular}{| p{1.4cm} | p{1.3cm} p{1.3cm} p{1.3cm} p{1.3cm} |}
                        \hline
                        Task & Cargo\_V1 & Cargo\_V2 & Cargo\_D6 & Cloud\\
                        \hline
                        Task\_V3   & \underline{\textbf{21}} & 25 & 31 & 61 \\
                        Task\_V4   & 25 & \underline{\textbf{23}} & 33 & 64 \\
                        Task\_V5   & 42 & 38 & \underline{\textbf{18}}   & 60 \\
                        \hline
                    \end{tabular}        
                }
                \vspace{1mm}
                \caption{Cargo selection}
                \vspace{-6mm}
                \label{tab:cargoSelect}
            \end{table}

        \subsubsection{Storage fault tolerance}
\begin{figure}[!h]
                \centering
                \vspace{-1mm}
                \includegraphics[scale=0.45]{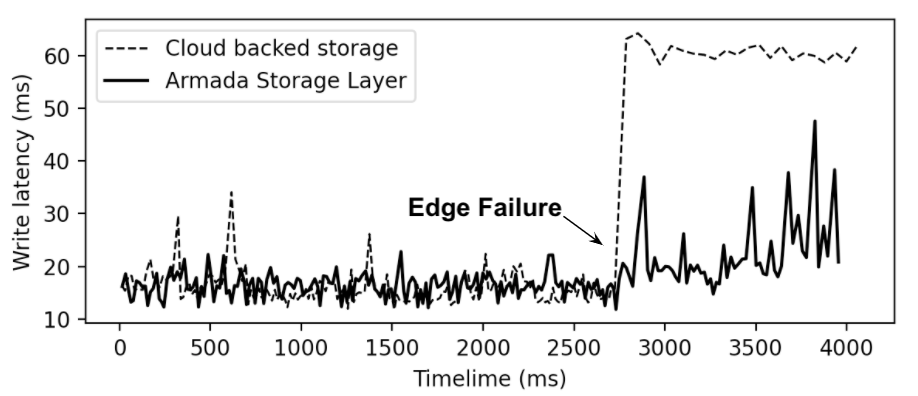}
                \vspace{-3mm}
                \caption{Continuous Cargo service on the edge}
                \vspace{-3mm}
                \label{fig:cargoSwitch}
                \end{figure}

            We demonstrate the storage fault tolerance behavior using the same experiment setup described in Section \ref{cargoSel}. In this experiment, we only focus on the read latency from Task\_V5’s perspective. Figure \ref{fig:cargoSwitch} shows that Task\_V5 can immediately switch to the Cargo\_V2 upon Cargo\_D6 failure. Thus, the Armada storage layer can guarantee continuous low-latency data access for edge services compared to a Cloud-backup scenario. This experiment also shows the benefits of exploiting volunteer resources when dedicated edge resources are not available.

        \subsubsection{Effect of Consistency}
\begin{figure*}[t]
                    \centering
                    \includegraphics[scale=0.47]{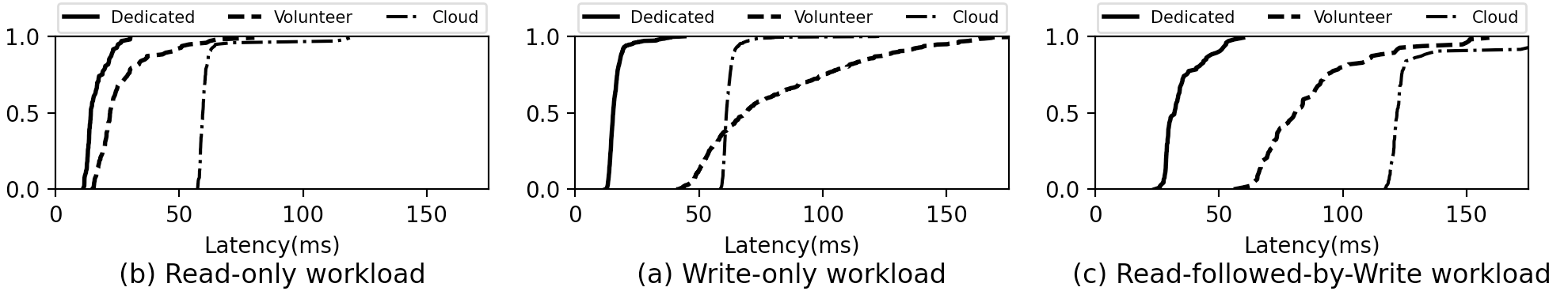}
                    \vspace{-3mm}
                    \caption{Read-Write latency for Strong Consistency}
                    \vspace{-3mm}
                    \label{fig:5-5-1}
                \end{figure*}

\begin{figure*}[t]
                    \centering
                    \includegraphics[scale=0.47]{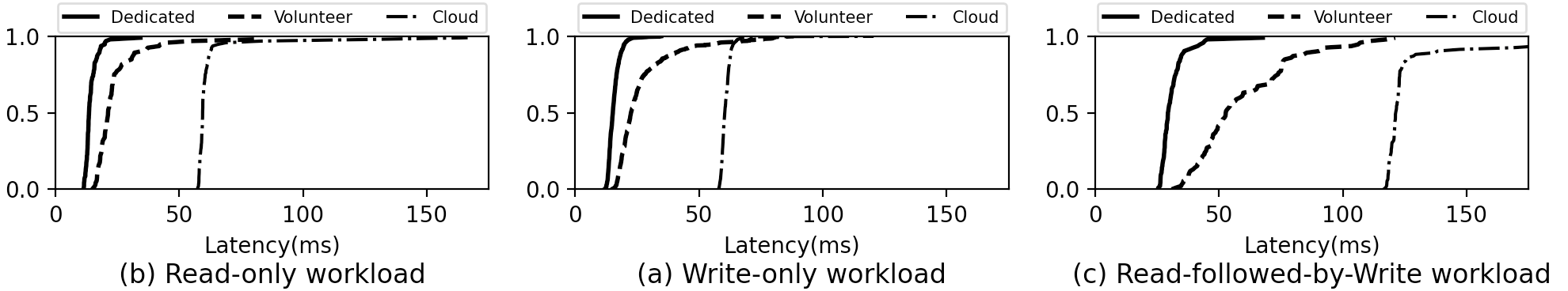}
                    \vspace{-3mm}
                    \caption{Read-Write latency for Eventual Consistency}
                    \vspace{-4mm}
                    \label{fig:5-5-2}
                \end{figure*}

            We run three workloads to explore the effect of different consistency policies in Armada. We also separate the performance for dedicated, volunteer edge resources and cloud to illustrate the benefits of exploiting volunteer resources for edge storage.
            We set up three configurations using dedicated Cargos, volunteer Cargos, and Cloud-located Cargos for both strong and eventual consistency scenarios. All edge nodes and users are loosely coupled with each other in real-world heterogeneous environments. As shown in Figure \ref{fig:5-5-1} and Figure \ref{fig:5-5-2}, we record the data I/O latency with varying configurations, consistency policies, and workload types.

            Figures \ref{fig:5-5-1} (a) and \ref{fig:5-5-2} (a) show that strong and eventual consistency have similar read latency since no data propagation is required for the read-only workload.
            Figures \ref{fig:5-5-1} (b) and \ref{fig:5-5-2} (b) show that the strong consistency for volunteer Cargos can cause higher latency than the cloud since volunteer nodes are loosely coupled, leading to high data propagation overhead.
            Similar to write-only workload, Figures \ref{fig:5-5-1} (c) and \ref{fig:5-5-1} (c) show that strong consistency has higher overhead caused by synchronized data propagation. Based on the above, volunteer Cargos in Armada exhibit similar performance compared to dedicated Cargos using eventual consistency. It also demonstrates the benefits of utilizing volunteer edge storage over the cloud for low-latency data access.

\section{Related Work}
	Several different research projects investigate the utilization of volunteer resources for both compute and storage \cite{bib:boinc, bib:bittorrent, bib:volunteersurvey}. Nebula \cite{ryden2014nebula} is a geo-distributed edge cloud that uses volunteer on an otherwise dedicated resource system to carry out data-intensive computing infrastructure for intensive computation and data storage with a NaCI sandbox. The NaCl sandbox is limited memory space and computation which defers it from running compute-intensive applications. Ad Hoc Cloud System \cite{mcgilvary2015ad} and cuCloud \cite{mengistu2018cucloud} are volunteer systems that harvest resources from sporadically available volunteer nodes, however, they lack locality or performance-aware mechanisms. Some groups have investigated running compute-intensive tasks on edge nodes based on MapReduce \cite{carson2019mandrake, costa2012large}. These studies aim to handle resource allocation and data durability, however they are mainly designed for heavy computation with less concern about data storage. In industry, K3s \cite{bib:k3s} is a lightweight version of kubernetes \cite{brewer2015kubernetes}, specifically designed for edge or IoT scenarios. KubeEdge \cite{kubeedge} leverage computing resources from the cloud and edge to coordinate both environments. However, they are still oriented to central clusters management without optimization on heterogeneous resources and locality.

	Storage at the edge can be categorized into offload (offload data to edge and sync with cloud), aggregate (Data collected from multiple devices to the edge) and P2P (data generated by one device shared with another) \cite{bib:fogStoreModels1, bib:fogStoreModels2}. Most of the existing storage systems focuses on offload and aggregate models. P2P storage is not explored much due to concerns of data security and synchronization difficulties across unreliable devices. CloudPath \cite{bib:cloudpath} uses PathStore \cite{bib:pathStore}, an eventually consistent datastore with persistent data on cloud and partial replicas on edge. The store may have a degraded performance when new data is queried frequently. SessionStore \cite{bib:sessionStore} is a hierarchical datastore that guarantees session consistency using session-aware reconciliation algorithms built on top of Cassandra \cite{bib:cassandra} and hence support client mobility to an extend. DataFog \cite{bib:datafog} is an IoT data management infrastructure which places replica based on spatial locality, addresses sudden surges in demand using a location-aware load balancing policy and evicts and compresses data based on temporal relevance. However, it does not support network proximity based node selection. FogStore \cite{bib:fogstore} is a geo-distributed key-value infrastructure that places replicas based on latency of data access. Also, to ensure fault tolerance similar to DataFog, one of the replicas is kept at a remote location in FogStore. However, it does not take into account the limited storage capacities of heterogeneous storage nodes.

    \section{Conclusion}
	We presented the design of Armada, a densely distributed edge cloud infrastructure running on dedicated and volunteer resources. The lightweight Armada architecture and system optimization techniques were described, including performance-aware edge selection, auto-scaling and load balancing on the edge, fault tolerance, and in-situ data access. We illustrated how Armada served geo-distributed users in heterogeneous environments. An evaluation was performed in both real-world volunteer environments and emulated platforms. Compared to the locality-based approach and dedicated-resource-only scenario, Armada shows a 32\% - 52\% reduction in average end-to-end latency. We will formulate a service/data placement problem to identify suitable nodes for deploying services and storing data for the next step. We also plan to carry out an online churn analysis to quantify the volunteer node stability, which will play an essential part in the placement process. Furthermore, we will also explore different policies like service/data migration and dynamic replication with fine-grained consistency to support mobility in the future Armada version.

\bibliographystyle{unsrtnat}
\bibliography{ms}

\end{document}